\documentclass[runningheads]{llncs}
\usepackage[T1]{fontenc}
\usepackage{graphicx}
\usepackage{bm} % for bold math symbols (incl. Greek letters) with \bm{}
\usepackage{cite}
\usepackage{amsmath,amssymb,amsfonts}
\usepackage{algorithmic}
\usepackage{graphicx}
\usepackage{textcomp}
\usepackage{pifont}
\usepackage{xcolor}
\usepackage{graphicx}
\usepackage{wasysym}
\usepackage{flushend} % to balance columns on last page
\usepackage{multirow}
\usepackage[normalem]{ulem}
\useunder{\uline}{\ul}{}

\usepackage{bm}
\usepackage{comment}
\usepackage{makecell}
\usepackage{tabularx}
\usepackage{graphicx}
\usepackage{multirow}
\usepackage[hidelinks]{hyperref}

\usepackage{booktabs}

\usepackage[english]{babel}
\addto\extrasenglish{}
\addto\extrasenglish{}
\addto\extrasenglish{}
\addto\extrasenglish{}
\addto\extrasenglish{}
\addto\extrasenglish{}

\newcolumntype{A}{>{\raggedright\arraybackslash}p{0.8cm}}  % Model column (left-aligned)
\newcolumntype{B}{>{\raggedright\arraybackslash}p{3.0cm}}  % Data column (left-aligned, widest content here)
\newcolumntype{C}{>{\centering\arraybackslash}p{1.2cm}}    % Numeric columns (centered)

\usepackage{acronym} 
\acrodef{AR}    {augmented reality}
\acrodef{ASD}   {audiovisual active speaker detection}
\acrodef{AVD}   {audiovisual diarisation}
\acrodef{CP}    {conversation Participant}
\acrodef{DeiT}  {data-efficient image transformer}
\acrodef{FIQA}  {Face Image Quality Assessment}
\acrodef{FVA}   {face-voice association}
\acrodef{FoV}   {field of view}
\acrodef{GRU}   {gated recurrent unit}
\acrodef{mAP}   {mean Average Precision}
\acrodef{MFCC}  {Mel frequency cepstral coefficient}
\acrodef{MLP}   {multilayer perceptron}
\acrodef{MTCNN} {Multi-task Cascaded Convolutional Neural Network}
\acrodef{SCAN}   {speaker comparison auxiliary network} 
\acrodef{SNR}   {signal-to-noise-ratio}
\acrodef{SoTA}  {state-of-the-art}
\acrodef{SL-ASD} {Self-Lifting for Audiovisual Active Speaker Detection}
\acrodef{VAD}   {voice activity detector}
\acrodef{V-TCN} {visual temporal convolutional network}

\begin{document}
\title{Ensembling Synchronisation-Based and Face–Voice Association Paradigms for Robust Active Speaker Detection in Egocentric Recordings}
\titlerunning{Robust ASD in Egocentric Recordings}
\author{Jason Clarke\inst{1} \and
Yoshihiko Gotoh\inst{1} \and
Stefan Goetze\inst{1,}\inst{2}}
\authorrunning{J.~Clarke et al.}
\institute{Speech and Hearing (SPandH), School of Computer Science, The University of Sheffield, UK, \email{\{jclarke8, y.gotoh, s.goetze\}@sheffield.ac.uk} 
\and
South Westphalia University of Applied Sciences, Iserlohn, Germany
\email{goetze.stefan@fh-swf.de}
}
\maketitle              % typeset the header of the contribution
\begin{abstract}
    \Ac{ASD} in egocentric recordings is challenged by frequent occlusions, motion blur, and audio interference, which undermine the discernability of temporal synchrony between lip movement and speech. Traditional synchronisation-based systems perform well under clean conditions but degrade sharply in first-person recordings. Conversely, \ac{FVA}-based methods forgo synchronisation modelling in favour of cross-modal biometric matching, exhibiting robustness to transient visual corruption but suffering when overlapping speech or front-end segmentation errors occur. In this paper, a simple yet effective ensemble approach is proposed to fuse synchronisation-dependent and synchronisation-agnostic model outputs via weighted averaging, thereby harnessing complementary cues without introducing complex fusion architectures. A refined preprocessing pipeline for the \ac{FVA}-based component is also introduced to optimise ensemble integration. Experiments on the Ego4D-AVD validation set demonstrate that the ensemble attains $70.2$\% and $66.7$\% \ac{mAP} with TalkNet and Light-ASD backbones, respectively. A qualitative analysis stratified by face image quality and utterance masking prevalence further substantiates the complementary strengths of each component. 
    % All code and model weights will be made available on GitHub.

\end{abstract}

\section{Introduction}
\label{sec:intro}

\Acf{ASD} involves identifying the framewise speaking activity of a candidate speaker through the joint analysis of audio signals and temporally aligned face tracks\cite{activespeakersincontext, ava-as, asdtransformer, buffy, talknet, Liao_2023_CVPR,ASDNet}. Traditional \ac{ASD} systems rely on modelling the temporal correspondence between speech in the audio signal and visual speech-related cues—such as lip movement or cheek posture\cite{asdtransformer}—in the candidate speaker’s face track. These synchronisation-based approaches assume audiovisual alignment as a prerequisite for detecting speech activity; this assumption dominates modern methods\cite{SPELL, LoCoNet, talknet, Liao_2023_CVPR}. Extensions to this framework incorporate contextual cues pertaining to inter-speaker relationships\cite{LeonAlcazar2021MAASMA, SPELL} and latent information describing the audible context of each scene\cite{clarke23-ASD-ASRU}, these extensions help to address multi-talker scenarios and environmental noise, respectively. However, such methods remain fundamentally contingent on the discernibility of audiovisual synchrony, resulting in these approaches still being vulnerable to the challenges posed in egocentric settings\cite{Ego4D, clarke2025speakerembeddinginformedaudiovisual}.

In egocentric recordings, e.g.~captured by head-worn recording devices, such as smart or \ac{AR} glasses, synchronisation-based \ac{ASD} performance deteriorates significantly when compared to their performance on exocentric benchmarks\cite{ava-as}. This is largely attributed to the prevalence of visual occlusions, motion blur, and audio interference from overlapping speech or environmental noise \cite{clarke23-ASD-ASRU, Ego4D, LoCoNet, clarke2025speakerembeddinginformedaudiovisual, 10888166, spellego4dchallenge}, all of which are common challenges in egocentric data. Since sychronisation-based methods require sustained discernable audiovisual cues, these challenges significantly degrade their performance.

% These challenges degrade the reliability of both modalities, as synchronisation-based methods require sustained, high-quality audiovisual cues. 

To circumvent these limitations, recent work by the authors of this paper has explored using \acf{FVA} for the task of \ac{ASD}, as exemplified by the \ac{SL-ASD} architecture~\cite{clarke2025EUSIPCO_FaceVoiceAssociatuon}. Generally, \ac{FVA}~\cite{seeking_shape_of_sound, Self-Lifting, disentangled_cross-modal_biometric_matching, single-branch} concerns the task of attributing pre-segmented speaker-invariant utterances to visible identities using cross-modal biometric information rather than temporal alignment. The \ac{SL-ASD} architecture~\cite{clarke2025EUSIPCO_FaceVoiceAssociatuon} builds upon this concept by adapting \ac{FVA}~\cite{Self-Lifting} for \ac{ASD}. This type of approach identifies and leverages transient high-quality facial frames to establish robust voice-face mappings, bypassing the need for fine-grained audiovisual cues being consistently discernable. Prior work~\cite{clarke2025EUSIPCO_FaceVoiceAssociatuon} has demonstrated robust performance in the context of egocentric recordings achieving \ac{mAP} scores close to the state-of-the-art despite using significantly less learnable parameters, exclusively for the task of \ac{ASD}. However, it has been observed~\cite{clarke2025EUSIPCO_FaceVoiceAssociatuon} that solely relying on face-voice associations introduces two main limitations: face-voice associations falter during speaker-variant utterances (i.e.~overlapping speech), and missed speech detections by the speaker-invariant front-end are harshly penalised when the pipeline is evaluated for \ac{ASD}, holistically. These shortfalls are distinct to the limitations of synchronisation-based methods which struggle more with visual degradation but typically have good recall when the speech signal is audible~\cite{TS-talknet, clarke2025speakerembeddinginformedaudiovisual, clarke23-ASD-ASRU}.  By leveraging the complementary strengths of these two paradigms, this work extends the existing \ac{SL-ASD} approach~\cite{clarke2025EUSIPCO_FaceVoiceAssociatuon} and proposes a simple yet effective ensemble approach that combines the benefits of synchronisation-agnostic (i.e. \ac{FVA}-based) and synchronisation-dependent methods of \ac{ASD}.

More precisely, the presented system integrates two symbiotic components as an ensemble: (i) a synchronisation-based model that captures temporal audiovisual correspondence \cite{talknet, Liao_2023_CVPR}, and (ii) a speaker-invariant segmentation front-end paired with a \ac{FVA} module, derived from prior work~\cite{clarke2025EUSIPCO_FaceVoiceAssociatuon} but with refinements for enhanced ensemble performance. The proposed ensemble aggregates output probability sequences from both systems, via weighted averaging, which mitigates each component's divergent failure modes. Although the ensemble mechanism is architecturally lightweight—requiring only a weighted mean fusion of two probability streams—its empirical efficacy demonstrates that synergistic modality insights can be leveraged without complex cross-model attention or gating networks. This simplicity encourages easier deployment on resource-constrained wearable devices.

% The synchronisation component excels in clean audiovisual conditions but degrades under modality-specific distortions, while the face-voice association module remains robust to occasional visual occlusions and transient audio interference. By combining these approaches, the ensemble mitigates the limitations of individual models, particularly in egocentric contexts. 

\newenvironment{tight_enumerate}{
\begin{enumerate}
  \setlength{\itemsep}{0pt}
  \setlength{\parskip}{0pt}
  \setlength{\topsep}{0pt}
}{\end{enumerate}}
\begin{tight_enumerate}
\item[] \hspace*{-3ex}\textbf{Contributions:} 
\item A lightweight late-fusion ensemble method for \ac{ASD} that combines syn\-chro\-ni\-sa\-tion-based and \ac{FVA}-based models, improving robustness under visual occlusion and audible noise.

% leverages complementary information extracted by synchronisation-based models and \ac{FVA}-based models to help disambiguate audible scenes with audible noise and visual degradation. 
\item A refined preprocessing pipeline for \ac{SL-ASD} to optimise ensemble performance.
\item Empirical validation on Ego4D-AVD: the ensemble achieves $70.2$\% and $66.7$\% \ac{mAP} for two synchronisation-based components (TalkNet and Light-ASD), marking a new state-of-the-art in the domain of egocentric \ac{ASD}.
% featuring more accurate label attribution
% \item Experiments on Ego4D-AVD involving two different synchronisation-based baselines (TalkNet and Light-ASD) which demonstrate the efficacy of the proposed ensemble, yielding $70.2$\% and $66.7$\% \ac{mAP}, respectively.
\item Qualitative analysis of performance including granular evaluations stratified by \ac{FIQA} and randomised utterance masking prevalence to demonstrate the vulnerabilities and strengths of each component of the ensemble.
\end{tight_enumerate}

\section{Methodology}
\label{sec:method}
This section first provides a brief overview of the typical single-candidate syn\-chro\-ni\-sa\-tion-based paradigm used for \ac{ASD} in~\autoref{ssec:sb_asd}, and then describes the \ac{FVA}-based approach to \ac{ASD} used by this work in~\autoref{ssec:fva_asd}. Finally, the details of the proposed ensemble method, which effectively combines the two synergistic approaches, are presented in \autoref{ssec:ensemble}.

\subsection{Synchronisation-based Approach to Active Speaker Detection}
\label{ssec:sb_asd}
\begin{figure}[!htb]
  \centering
    \includegraphics[scale=0.3]{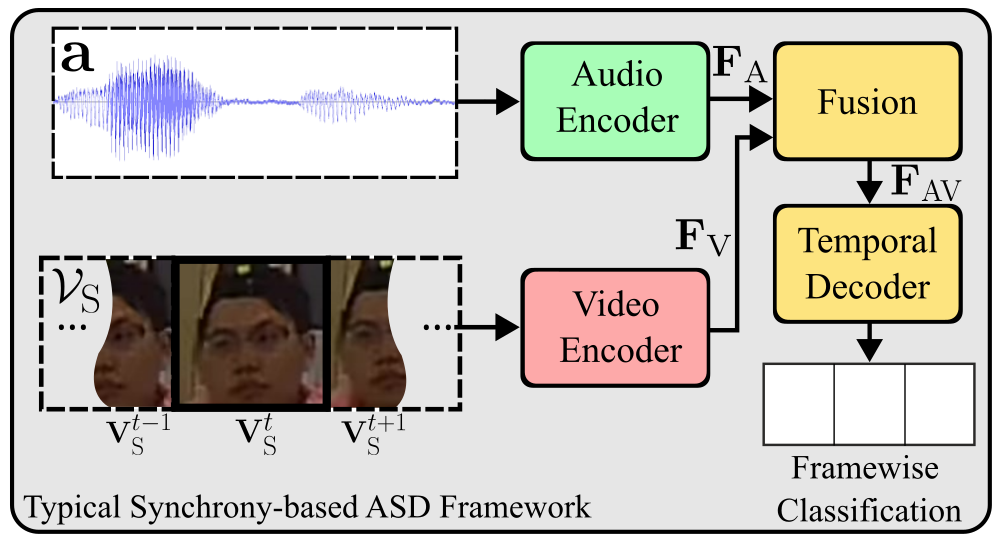}
  \caption{Typical synchronisation-based single-candidate approach to \ac{ASD}\cite{talknet, Liao_2023_CVPR}.}
\label{fig:sync_asd}
\end{figure}

Conventional single-candidate \ac{ASD} systems operate by assessing the temporal alignment between cues indicative of speech in a given face track signal and the concurrent audio signal as illustrated in~\autoref{fig:sync_asd}. 

A face track $\mathcal{V}_{S} = \{\mathbf{V}_{\mathrm{S},1}, \dots, \mathbf{V}_{\mathrm{S},T}\}$ is defined as a sequence of $T$ contiguous bounding box face crops $\mathbf{V}_{\mathrm{S},t}\in\mathbb{R}^{H\times W}$ of height $H$ and width $W$, centred on a single candidate speaker $\mathrm{S}$ and the concurrent audio signal is defined as a vector of $T_{\mathrm{A}}$ waveform samples $\mathbf{a}\in\mathbb{R}^{T_{\mathrm{A}}}$ (note that $T_{\mathrm{A}}$ differs from $T$ due to frame rate differences in audio and video modalities). 

First, an audio encoder processes the audio signal $\mathbf{a}$, and a video encoder processes face tracks $\mathcal{V}_S$, each producing an embedding with shared dimensions. Specifically, the audio branch yields $\mathbf{F}_{\mathrm{A}} \in \mathbb{R}^{T\times d}$ and the video branch yields $\mathbf{F}_{\mathrm{V}} \in \mathbb{R}^{T\times d}$, where $d$ is the embedding dimension of the respective encoders. These two embeddings are then fused to create a single multimodal representation $\mathbf{F}_{\mathrm{AV}}$.  Common fusion operations include channel-wise concatenation, element-wise summation\cite{Liao_2023_CVPR}, or attention-based weighting\cite{talknet}. Regardless, $\mathbf{F}_{\mathrm{AV}}$ encodes both audio and visual information at each video-frame.

Finally, a temporal decoder (e.g. a lightweight transformer or temporal convolutional network) is applied along the $T$ dimension of $\mathbf{F}_{\mathrm{AV}}$ to model longer-range dependencies in speech activity. A frame-wise classification head then produces probabilities indicating whether the candidate is active at each video-frame.  
% During training, a binary cross-entropy loss is minimized against ground-truth speaking annotations. 
This pipeline—embodied by architectures such as TalkNet\cite{talknet} and Light-ASD\cite{Liao_2023_CVPR}—relies fundamentally on audiovisual synchrony, requiring high-quality lip motion and clean audio to be consistently available for accurate detection.

\subsection{Face-Voice Association for Active Speaker Detection}
\label{ssec:fva_asd}

The face–voice association approach to \ac{ASD} replaces the need for explicit audiovisual synchronisation-based modelling by leveraging cross-modal biometric correspondence. This paper follows prior work, specifically the \ac{SL-ASD} architecture\cite{clarke2025EUSIPCO_FaceVoiceAssociatuon}, but deviates in terms of preprocessing implementation which has been optimised by this study for the ensemble approach described in \autoref{ssec:ensemble}. Hence, the method proposed here will be denoted as $\ac{SL-ASD}\dagger$, which is outlined as follows.

\subsubsection{Front-end Segmentation and Embedding}
Let $\mathcal{C}$ denote the set of video clips in a given dataset.
First, an off-the-shelf speaker-diarisation front-end~\cite{pyannote} is applied to the audio signal $\mathbf{a}_c$ of each clip $c$, segmenting each clip into a set of speaker-invariant utterances. Each utterance $\mathbf{u}_{c,i}$ is then embedded by a pretrained speaker-recognition model\cite{ecapa-tdnn} yielding an embedding $\mathbf{u}'_{c,i}\in\mathbb{R}^{d_{\mathrm{S}}}$ for all utterances, where $d_{\mathrm{S}}$ is the embedding dimension of the speaker recognition model. Collectively, these embeddings form
$\mathcal{U}' = \bigl\{\mathbf{u}'_{c,i}\mid c\in\mathcal{C},\;i=1,\dots,N_c\bigr\}$, where $N_c$ is the number of utterances in clip $c$. For this segmentation, the Pyannote Audio diarisation model~\cite{pyannote} is used because of its robust performance in the task of audio-only diarisation\cite{pyannote_ego4d_2022_diarisation_challenge}. 

Additionally, every face-crop image $\mathbf{V}_{\mathrm{S},T}$ in the dataset is embedded by a pretrained face-recognition model\cite{inceptionv1} yielding a hierarchical set of face-recognition embeddings 
$\mathcal{X} = \bigl\{ 
%\mathbf{f}_{c,s,t} \bigr)_{t=1}^{T_{c,s}} 
\mathbf{X}_{c,s} \mid s \in \mathcal{S}_c,\ c \in \mathcal{C} \bigr\}$,  
where each matrix \linebreak $\mathbf{X}_{c,s} = [\mathbf{x}_{c,s,1}, \mathbf{x}_{c,s,1}, ..., \mathbf{x}_{c,s,T_{c,s}}]$ contains face embedding vectors per speaker $s$ in clip $c$ and different $\mathbf{X}_{c,s}$ may be of different size due to variability of frames $T_{c,s}$ per clip and speaker. $\mathcal{S}_c$ is the set of visible identities in clip $c$, and $T_{c,s}$ is the number of frames for identity $s$ in clip $c$.  
%$\bigl( \mathbf{f}_{c,s,t} \bigr)_{t=1}^{T_{c,s}}$ 

%\begin{equation}
%    \mathbf{F}_{c,s} = [\mathbf{f}_{c,s,1}, \mathbf{f}_{c,s,1}, ..., \mathbf{f}_{c,s,T_{c,s}}]
%\end{equation}

\subsubsection{Self-Lifting for Active Speaker Detection}

During training, the audio component of each batch consists of several speaker-embeddings $\mathbf{u}'_{c,i} $  sampled from $\mathcal{U}'$ ensuring each utterance was taken from the same clip and spoken by the same identity (as per groundtruth annotation). During inference, since groundtruth annotation for utterance identity is not available, the audio component of each batch is simply a single speaker embedding. For both training and inference, the visual component of each batch comprises $\bigl\{ \mathbf{X}_{c,s} \mid \forall s \in \mathcal{S}_c, \bigr\}$, where $c$ refers to the clip from which the speaker embedding(s) in the audio component of the batch were taken from. Each component of the batch is then fed through the relevant branch of the pretrained Self-Lifting\cite{clarke2025EUSIPCO_FaceVoiceAssociatuon} model, resulting in $\mathbf{U}'' \in \mathbb{R}^{N_u \times d}$ and $\mathbf{X}'_c \in \mathbb{R}^{|\mathcal{S}_c| \times (\max_{s \in \mathcal{S}_c} T_{c,s}) \times d}$, from the audio and visual branches, respectively. Here, $N_u$ denotes the number of utterances in the audio component of the batch, which is set to $1$ during inference.
To account for variable visual quality—common in egocentric footage—a lightweight transformer encoder is applied over each sequence dimension (frame dimension) for each visible identity in $\mathbf{X}'_c$.  Through its self-attention mechanism, low-quality frames (e.g. blurred or occluded) are down-weighted, and the resulting sequence is mean-pooled to produce a single quality-aware face-recognition embedding for each identity in the visible component of the batch, resulting in $\mathbf{X}''_c \in \mathbb{R}^{|\mathcal{S}_c| \times 1 \times d}$.  

\begin{figure}[!htb]
  \centering
    \includegraphics[scale=0.55]{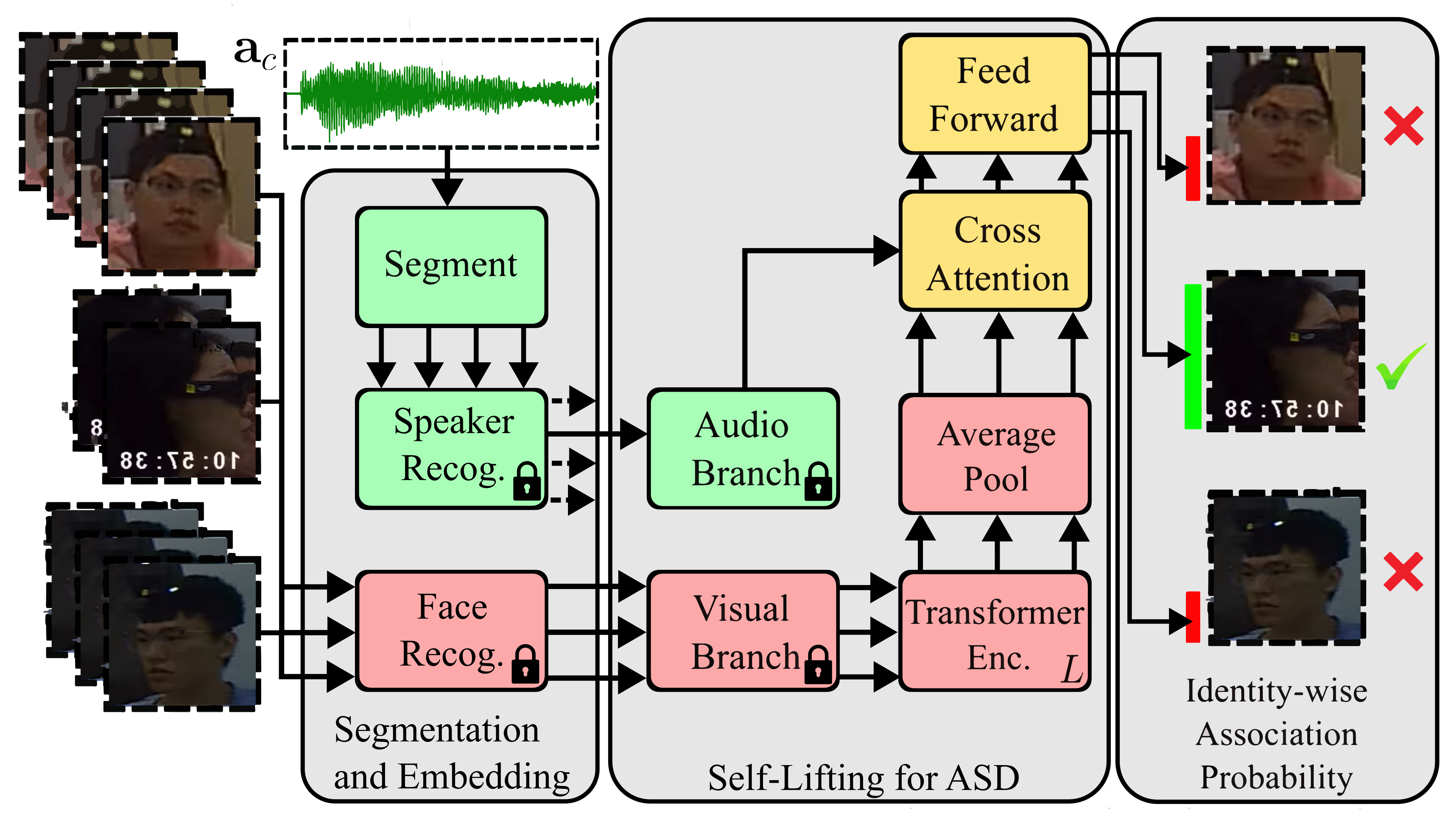}
  \caption{$\ac{SL-ASD}\dagger$ framework. Colours indicate modality. Bars adjacent to faces on the right indicate probability of a face-voice match.}
\label{fig:SL_ASD}
\end{figure}

Finally, cross-modal association scores are computed by measuring similarity between the embedded utterance and each aggregated face-recognition embedding in the video component of the processed batch, as illustrated in \autoref{fig:SL_ASD}. Specifically, scaled dot-product cross-attention is used to produce a matching probability that a given utterance was spoken by each visible identity. This pure face–voice association pipeline thus attributes each speech segment to the most likely visible identity, relying only on biometric consistency rather than audiovisual synchrony.

% \begin{figure*}[!ht]
%   \centering
%     \includegraphics[scale=0.5]{figures/sl_ASD_v1.png}
%   \caption{\ac{SL-ASD} framework\cite{clarke2025EUSIPCO_FaceVoiceAssociatuon}, dotted lines of of speaker embeddings indicate only utterances belonging to a single speaker are passed through the pipeline at a time during training. Colours indicate modality. Bars adjacent to end faces indicate probability of a face-voice match.}
% \label{fig:SL_ASD}
% \end{figure*}

\subsection{Ensembling Synchronisation-Based and \ac{FVA}-based Approaches to Audiovisual Active Speaker Detection}
\label{ssec:ensemble}

While synchronisation-based (cf.~\autoref{ssec:sb_asd}) and \ac{FVA}-based approaches (cf.~\autoref{ssec:fva_asd}) offer complementary strengths, each exhibits vulnerabilities under challenging audiovisual conditions when used in isolation. To mitigate these limitations, an ensemble strategy is employed which fuses predictions from both paradigms by averaging their respective probability sequences.

Let $\mathbf{p}_{\mathrm{sync}} \in [0,1]^T$ denote the frame-level speaking probabilities predicted by a synchronisation-based model for a given face track. Let $\mathbf{p}_{\mathrm{assoc}} \in [0,1]^T$ denote the probability sequence derived from the \ac{FVA}-based model for the same hypothesis track as in $\mathbf{p}_{\mathrm{sync}}$. Forming $\mathbf{p}_{\mathrm{assoc}}$ is acheived by projecting the face-voice matching probability uniformly across all concurrent frames in each face track that temporally overlaps with the given utterance. The final ensemble prediction $\mathbf{p}_{\mathrm{ens}}$ is then computed via framewise weighted mean averaging, where $\alpha$ is a mixing coefficient determined empirically:
\begin{equation}
  \mathbf{p}_{\mathrm{ens}}
  = \alpha\,\mathbf{p}_{\mathrm{sync}}
    + (1 - \alpha)\,\mathbf{p}_{\mathrm{assoc}}.
\end{equation}
This late-fusion scheme requires no additional training and yields a probability sequence that integrates both dynamic synchronisation cues and cross-modal biometric consistency. The resulting ensemble consistently outperforms either constituent method when used in isolation, particularly in scenarios with degraded visual quality or non-frontal faces (cf.\autoref{ssec:sota}).

% where either synchrony-based or biometric-based approaches may underperform in isolation.

\section{Experiments}
\label{sec:exps}

This section briefly introduces the egocentric Ego4D %and the exocentric AVA 
dataset used in this work in \autoref{ssec:datasets}, the implementation details in \autoref{ssec:imp_details}, and the evaluation metrics used throughout in \autoref{ssec:eval_metrics}.

\subsection{Ego4D Dataset for Egocentric Audiovisual Diarisation}
\label{ssec:datasets}

%\subsubsection{Ego4D}
The Ego4D dataset~\cite{Ego4D} comprises egocentric video recordings, totalling $572$ unique clips each lasting five minutes in duration, some of which were captured simultaneously. The data was obtained using various wearable devices using $1080$p video. The audio signals are standardised to a single-channel in $16$~kHz format. Video-frames were recorded at $30$~Hz. The dataset reflects real‐world conditions -- featuring fluctuating lighting, frequent occlusions, and continuously changing viewpoints -- making it a particularly demanding testing scenario for \ac{ASD}. Ego4D-AVD is divided into three non‐overlapping folds: $379$ clips for training, $50$ for validation, and $133$ for testing. Because test labels are withheld, the original training set was further split by this work into $110$ clips for model training and $23$ for development, preserving the reserved validation set for final evaluation. Splits were created to ensure that no individual appears in more than one fold. 
% Within each clip, consistent inter‐track identity annotations are provided, allowing utterance–face matching to leverage frames across different tracks; without these annotations, comparisons would be restricted to frames within the same track alone. ```

%\subsubsection{AVA-ActiveSpeaker}
%AVA-ActiveSpeaker~\cite{ava-as} is the pioneering large-scale benchmark for \ac{ASD}, assembled from $262$ Hollywood film excerpts recorded from an exocentric viewpoint. The dataset is partitioned into $120$ training, $33$ validation, and $109$ test clips, with per-frame labels marking speaking activity and yielding over $5.3$~million annotated face bounding boxes.  
%Faces are grouped into temporally coherent tracks, each linked to a single identity, facilitating approximately $3.65$~million speech vs. non‐speech annotations spanning $38.5$~hours of footage. The recordings exhibit frequent occlusions, low-resolution face crops, background audio noise, and variable lighting conditions, providing an more-forgiving type of data than Ego4D but nevertheless still challenging.  

\subsection{Implementation Details}
\label{ssec:imp_details}
\subsubsection{Synchronisation-Based Models} For this component of the ensemble two different \ac{ASD} systems were used as baselines, namely TalkNet~\cite{talknet} and Light-ASD~\cite{Liao_2023_CVPR}. These architectures were implemented under the exact configurations and hyperparameters specified in their original manuscripts apart from the training duration. Each model was trained independently $10$ times for $30$ epochs, and the checkpoint achieving the best performance on the development-set of Ego4D was selected. Finally, the selected checkpoints were employed to generate the synchronisation-based predictions incorporated into the ensemble.
\subsubsection{Self-Lifting for Audiovisual Active Speaker Detection} The $\ac{SL-ASD}\dagger$ implementation was similar to that described in~\cite{clarke2025EUSIPCO_FaceVoiceAssociatuon}. Specifically, the front-end utterance segmentation was performed on a clipwise basis using the Pyannote.audio-speaker-diarization-3.1 system~\cite{pyannote} to extract speaker-invariant utterances. Speaker-recognition embeddings were obtained from these utterances using the ECAPA-TDNN~\cite{ecapa-tdnn} model, pretrained on VoxCeleb2~\cite{voxceleb2}. Face-recognition embeddings were extracted from all face-track frames in the dataset via Inception-V1~\cite{inceptionv1} pretrained on VGG-Face2~\cite{vggface2}. For finetuning of the Self-Lifting audio and video encoder branches, the model was instantiated with the implementation described in its original manuscript~\cite{Self-Lifting}, except the number of cluster centroids, which was reduced to $50$ to better reflect the number of distinct identities present in Ego4D. In the \ac{ASD} adaptation (\ac{SL-ASD}~\cite{clarke2025EUSIPCO_FaceVoiceAssociatuon}), all original framework parameters were frozen, and only the transformer encoder, the cross-attention module, and the feed-forward layer were trained explicitly for \ac{ASD} (cf.~\autoref{fig:SL_ASD}). During training, each batch’s audio component comprised all utterances for a single clipwise identity, while its video component included all face-track frames for every visible identity in the clip; during validation and inference, the audio component was limited to single utterances. Optimisation was carried out using Adam with an initial learning rate of $1\times10^{-5}$, decayed by a factor of $0.2$ every $5$ epochs, and a single transformer layer with four attention heads was employed for both the encoder and cross-attention.
\subsubsection{Face Quality Assessment}To perform a granular evaluation of the various approaches to \ac{ASD} considered by this work (cf.\ \autoref{ssec:qa}), a method to quantify the visual quality of the frames in each face‐track was employed. In analogy to the well‐established domain of \acf{FIQA}~\cite{unsupervised-FIQA,faceqnet,magface}, the per‐frame recognisability of the candidate speaker was inferred via the confidence score produced by the pretrained \ac{MTCNN} face detector~\cite{Zhang2016MTCNN}. Specifically, every cropped face image in a groundtruth track was passed through \ac{MTCNN}, and the resulting detection probabilities were recorded. These per‐frame scores were then averaged to yield a single, track‐level quality metric. 
% Finally, face‐tracks were stratified into discrete quality bins (e.g.\ low, medium, and high) based on their mean confidence, thereby enabling a stratified analysis of model performance under varying degrees of facial distortion, occlusion, and lighting conditions. This heuristic approach was chosen for its computational efficiency, its direct reliance on an off-the-shelf detector, and its demonstrated correlation with more complex, recognition‐aware quality measures.

\subsection{Evaluation Metrics}
\label{ssec:eval_metrics}
For holistic evaluation, each system is evaluated for \ac{ASD} using the Cartucho object detection \ac{mAP} metric~\cite{cartucho}, which is in alignment with the \ac{mAP} protocol established by the PASCAL VOC2012 challenge~\cite{pascal-voc-2012}. This evaluation strategy is consistent with the framework adopted by the Ego4D audiovisual diarization challenge~\cite{Ego4D} and is widely employed in recent \ac{ASD} research~\cite{clarke23-ASD-ASRU, clarke2025speakerembeddinginformedaudiovisual}. Owing to the absence of ground truth annotations for the test folds in Ego4D, all results are reported on its validation folds, in accordance with prevailing conventions in the literature~\cite{activespeakersincontext, ASDNet, clarke23-ASD-ASRU, SPELL, EASEE, LoCoNet}. The validation fold is exclusively reserved for testing purposes and are not used during model development. For the evaluations presented in \autoref{ssec:qa}, the problem is reformulated as a binary classification task, with metrics computed using the scikit-learn~\cite{scikit-learn} implementation of average precision.

\section{Results}
\label{sec:res}

\subsection{Comparison with State-of-the-Art Methods}
\label{ssec:sota}

To assess the efficacy of the proposed ensemble, its performance is evaluated holistically against leading \ac{ASD} systems.  \autoref{tab:sota} summarises \ac{mAP} and parameter counts for each method on the validation fold of the Ego4D-AVD benchmark.

\begin{table}[!htb]
\centering
\caption{Comparison with state-of-the-art \ac{ASD} systems on the validation fold of Ego4D.``\ac{ASD} Params. [M]" denotes the number of learnable-parameters each system uses exclusively for the task of \ac{ASD}. All values are taken from published literature except ensemble approaches. $\ac{SL-ASD}\dagger$ indicates the modified implementation of \ac{SL-ASD}~\cite{clarke2025EUSIPCO_FaceVoiceAssociatuon} used by this work.
}
\label{tab:sota}
%\resizebox{\columnwidth}{!}{%
    \begin{tabular}{cccc}
      \hline
      \textbf{Model} & \textbf{Ensemble} & \textbf{mAP [\%]} & \textbf{ASD Params. [M]} \\ 
      \hline
      TalkNet\cite{Ego4D}  & \ding{55} & 51.0 & 15.1 \\
      Light ASD\cite{clarke2025speakerembeddinginformedaudiovisual}& \ding{55} & 54.3 & 1.0  \\
      \ac{SL-ASD}\cite{clarke2025EUSIPCO_FaceVoiceAssociatuon}               & \ding{55} & 59.7 & \textbf{0.4} \\
      SPELL\cite{spellego4dchallenge}& \ding{55} & 60.7 & $\!>\!22.5$ \\
      LoCoNet\cite{LoCoNet}& \ding{55} & 68.4 & 33.5 \\
      \hline
      Light ASD + TalkNet  & \ding{51} & 64.1 & 16.1 \\
      Light ASD + $\ac{SL-ASD}\dagger$         & \ding{51} & 67.1 & 1.4  \\
      \textbf{TalkNet+$\ac{SL-ASD}\dagger$} & \ding{51} & \textbf{70.2} & 15.5 \\
      \hline
    \end{tabular}
%  }
\end{table}

When fused via weighted averaging, the synchronisation-based TalkNet model in conjunction with the face–voice association-based $\ac{SL-ASD}\dagger$ model yield a combined \ac{mAP} of $70.2$\%, outperforming both individual baselines (TalkNet: $51.0$\%; SL-ASD: $60.7$\%) by a substantial margin.  Crucially, this gain cannot be attributed merely to increased model capacity. This is illustrated by comparing the performance of an ensemble of two synchronisation-based approaches (TalkNet + Light-ASD) of $64.1$\% with that of Light-ASD + $\ac{SL-ASD}\dagger$ of $66.7$\%. While the former yields a significant improvement over its respective baselines, it still exhibits weaker performance than the latter, despite requiring significantly more learnable parameters. This indicates that combining synchronisation-based approaches with \ac{FVA}-based approaches leverages truly complementary cues.

Moreover, the TalkNet + $\ac{SL-ASD}\dagger$ ensemble establishes a new state of the art, surpassing the recent LoCoNet\cite{LoCoNet} model by $1.8$\% absolute \ac{mAP} while using fewer than half of its learnable parameters exclusively dedicated to \ac{ASD}. This demonstrates that simple late fusion of heterogeneous \ac{ASD} paradigms can yield superior accuracy-efficiency trade-offs compared to monolithic architectures, even those that effectively leverage contextual information.

\subsection{Qualitative Analysis}
\label{ssec:qa}

To further investigate the hypothesis that \ac{FVA}-based models leverage information complementary to that of synchronisation-based approaches, a stratified evaluation was conducted. Face tracks were grouped into discrete bins based on their average face quality scores, enabling a detailed analysis of model performance under varying degrees of visual degradation, including factors such as blur, occlusion, and suboptimal lighting conditions.

\begin{figure}[!htb]
  \centering
    \includegraphics[scale=0.2]{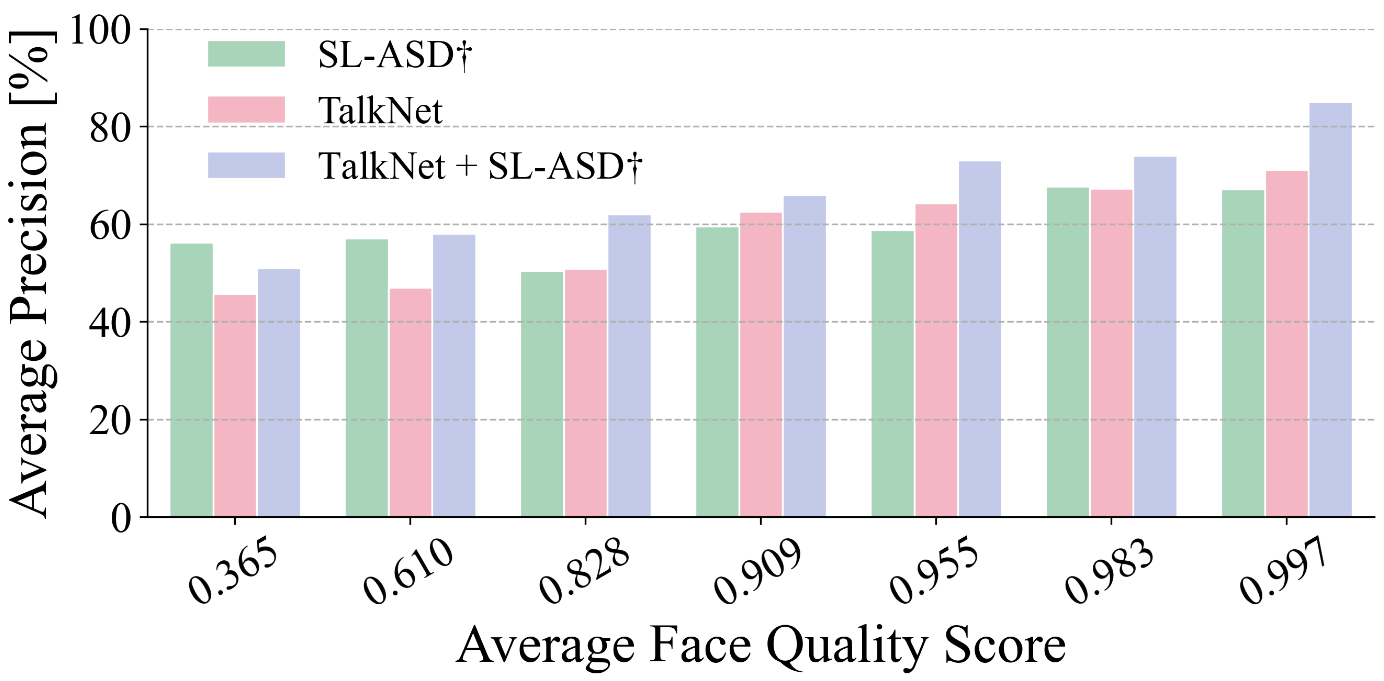}
  \caption{Comparison of synchronisation-based (TalkNet~\cite{talknet} pink bar), \ac{FVA}-based (\ac{SL-ASD}~\cite{clarke2025EUSIPCO_FaceVoiceAssociatuon}, green bar), and ensemble-based (blue bar) approaches to \ac{ASD}, evaluated on strata of equal size (each comprising tracks with similar average face quality scores). Lower face quality scores indicate tracks with greater visual distortion or occlusion. Irregular face quality score incrementation is due to a non-uniform distribution of trackwise visual quality.}
\label{fig:apvsfq}
\end{figure}

The results of this evaluation, shown in \autoref{fig:apvsfq}, reveal that synchronisation-based models, as speculated\cite{clarke23-ASD-ASRU, clarke2025EUSIPCO_FaceVoiceAssociatuon, clarke2025speakerembeddinginformedaudiovisual}, exhibit a significant decline in performance as face quality deteriorates. This sensitivity is attributed to their reliance on precise visual cues—particularly lip movements and cheek posture~\cite{asdtransformer}—that must be consistently discernible throughout the duration of the face track. In contrast, the \ac{FVA}-based model, $\ac{SL-ASD}\dagger$, demonstrates a more stable performance across all quality bins. Its robustness stems from the ability to identify and utilise even a limited number of high-quality frames within a sequence. The transformer encoder within $\ac{SL-ASD}\dagger$ effectively down-weights low-quality frames and emphasizes those that are most informative for identity recognition. This mechanism allows the model to maintain reliable speaker attribution despite transient visual distortions.

\begin{figure}[!htb]
  \centering
    \includegraphics[scale=0.2]{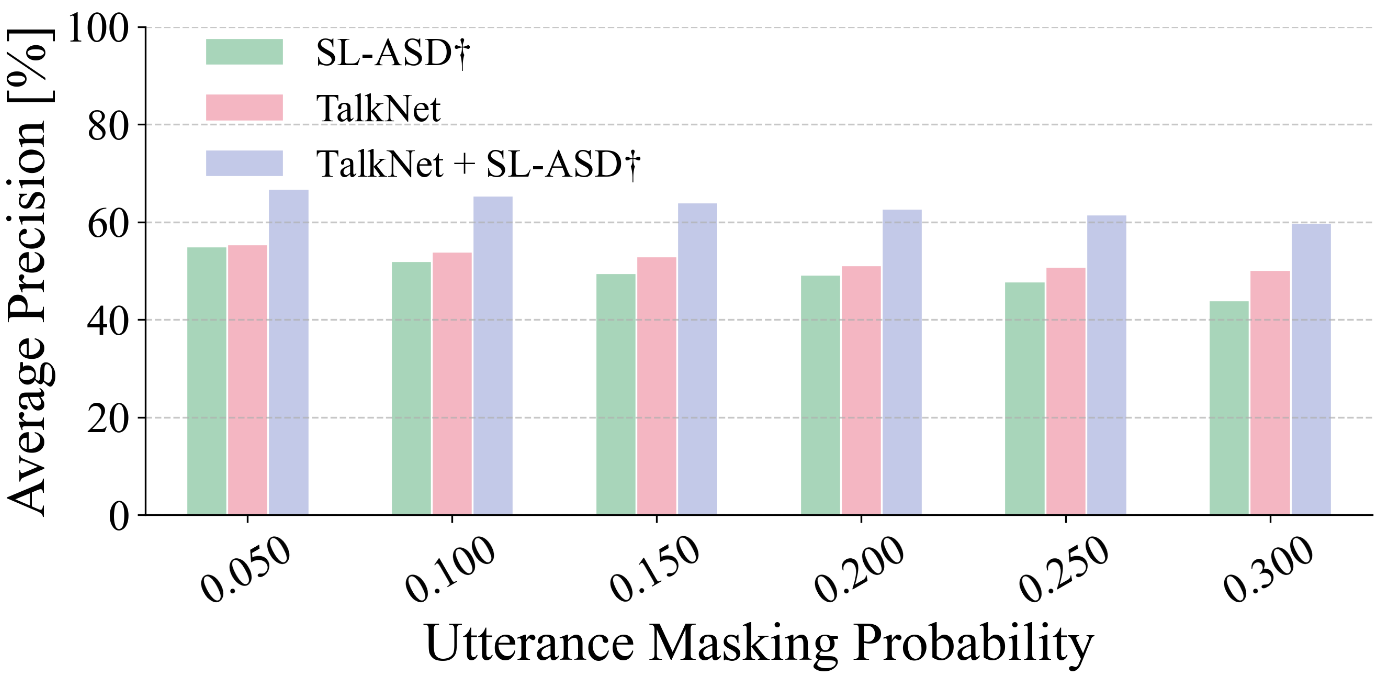}
  \caption{Comparison of three approaches to \ac{ASD} on the Ego4D validation set: synchronisation-based (TalkNet~\cite{talknet}, pink bar), \ac{FVA}-based (\ac{SL-ASD}~\cite{clarke2025EUSIPCO_FaceVoiceAssociatuon}, green bar), and ensemble-based (blue bar) methods. The evaluation is performed with randomised masking applied specifically to utterance regions within the audio signals, simulating various levels of audio signal degradation.}
\label{fig:apvsump}
\end{figure}

Conversely, \autoref{fig:apvsump} conveys the effect of audio degradation on each approach. As the probability of randomised utterance masking is increased, only a modest reduction in average precision is exhibited by the synchronisation-based model, owing to its ability to leverage cross-modal information, in this case video, when the audio is obscured. By contrast, a steeper decline is observed for the face–voice association–based $\ac{SL-ASD}\dagger$, since uninterrupted utterance segments are required by its speaker-invariant front-end for robust speaker embedding extraction. Crucially, higher overall performance across all masking levels is maintained by the ensemble approach, which leverages both streams to compensate for audio distortions that would otherwise impair face–voice association. These findings further substantiate that synchrony-dependent and synchrony-agnostic paradigms leverage complementary information for \ac{ASD}.

% \begin{figure}[!htb]
%   \centering
%     \includegraphics[scale=0.15]{figures/AP_vs_alpha.png}
%   \caption{Simulation to empirically determine the optimal value for $\alpha$ for each ensemble approach. Acquired via Ego4D-AVD development fold.}
% \label{fig:apvsalpha}
% \end{figure}

\section{Conclusion}
\label{sec:conc}

In this work, a lightweight late-fusion ensemble for \ac{ASD} was proposed, combining synchronisation-based and \ac{FVA}–based models to enhance robustness under visual occlusion and audio interference. The preprocessing pipeline of \ac{SL-ASD} was refined to optimise its integration within the ensemble, leading to consistent performance gains. Empirical validation on the Ego4D-AVD validation set demonstrated that the ensemble attains $70.2$\% and $66.7$\% \ac{mAP} when paired with TalkNet and Light-ASD backbones, respectively—establishing a new state-of-the-art in \ac{ASD}. Finally, a qualitative analysis stratified by face quality and utterance masking prevalence was conducted, revealing the complementary strengths and failure modes of each model component. Collectively, these findings substantiate that simple yet principled fusion of synchrony-dependent and synchrony-agnostic streams can reliably mitigate modality-specific degradations in challenging egocentric scenarios.  

\begin{credits}
\subsubsection{\ackname} 

This work was supported by the Centre for Doctoral Training in Speech and Language Technologies (SLT) and their Applications funded by UKRI [grant number EP/S023062/1]. This work was also funded in part by Meta.

% \subsubsection{\discintname}
% It is now necessary to declare any competing interests or to specifically
% state that the authors have no competing interests. 

\end{credits}
\bibliographystyle{splncs04}
\bibliography{refs25}




% LNCS Homepage, \url{http://www.springer.com/lncs}, last accessed 2023/10/25
% \end{thebibliography}

\end{document}